# Can we use Google Scholar to identify highly-cited documents?

Alberto Martin-Martin[1], Enrique Orduna-Malea[2], Anne-Wil Harzing[3], Emilio Delgado López-Cózar[1]*

[1] Facultad de Comunicación y Documentación, Universidad de Granada, Colegio Máximo de Cartuja s/n, 18071, Granada, Spain.
[2] Institute of Design and Manufacturing, Universitat Politècnica de València, Camino de Vera s/n, 46022, Valencia, Spain.
[3] Middlesex University Business School, NW4 4BT, London, UK.

* Corresponding author: E-mail address: edelgado@ugr.es (E. Delgado).

**Abstract**
The main objective of this paper is to empirically test whether the identification of highly-cited documents through Google Scholar is feasible and reliable. To this end, we carried out a longitudinal analysis (1950 to 2013), running a generic query (filtered only by year of publication) to minimise the effects of academic search engine optimisation. This gave us a final sample of 64,000 documents (1,000 per year). The strong correlation between a document's citations and its position in the search results ($r = -0.67$) led us to conclude that Google Scholar is able to identify highly-cited papers effectively. This, combined with Google Scholar's unique coverage (no restrictions on document type and source), makes the academic search engine an invaluable tool for bibliometric research relating to the identification of the most influential scientific documents. We find evidence, however, that Google Scholar ranks those documents whose language (or geographical web domain) matches with the user's interface language higher than could be expected based on citations. Nonetheless, this language effect and other factors related to the Google Scholar's operation, i.e. the proper identification of versions and the date of publication, only have an incidental impact. They do not compromise the ability of Google Scholar to identify the highly-cited papers.

## 1. Introduction

Google Scholar is a free academic web search engine that indexes scholarly literature across a wide array of disciplines, document types and languages (Ortega, 2014). It therefore specialises in finding and identifying bibliographic scholarly material, as well as providing a number of value-added services, such as direct access to the full texts (although for legal reasons this is not possible for all documents), the number of citations received by each document, and the number of different versions of the document. Google Scholar first appeared in November 2004, coinciding almost exactly with the launch of Scopus (*Reed Elsevier*, 2004). This meant that both products entered onto the market for bibliometric databases at the same time, a market in which up till then the Web of Science (WoS) had held a monopoly. The products had diametrically opposite approaches, however. Whereas Google Scholar was conceived as an open, dynamic but uncontrolled and fully automated product (Jacsó, 2005), Scopus positioned itself as a controlled product: with human intervention, closed, more static and designed to compete directly with WoS (Jacsó, 2008c).

Web of Science and Scopus began a rivalry as databases geared to the world of academic impact assessment. In the meantime, Google Scholar operated in another, complementary, sector: searching, locating and accessing academic information, in the broad sense of the term. Before the latter had completed its first year of operation, both *Science* (Leslie, 2004) and *Nature* (Butler, 2004) had already made mention of its impact on the scientific community.

Studies of Google Scholar as a scholarly information search tool have been undertaken primarily by the library sector. Three different phases can be discerned in these studies. At first, Google Scholar was observed with curiosity and scepticism. This phase was followed by a period of systematic study when it received harsh criticism. Finally, the third phase was one of optimism about its potential to reach 100% of the information available online for an institution, person, journal or other scholarly communication channel (Howland et al, 2009).

Among these studies, we can identify, on the one hand, those which look to understand how the quality and the usefulness of the product is perceived by different types of users, such as students (Carpenter, 2012), academics (Schonfeld & Housewright, 2010; Housewright et al, 2013; Van Noorden, 2014) and information professionals (Giles, 2005; Giustini & Barsky, 2005; Ettinger, 2008). On the other hand, there are the studies comparing the performance of Google Scholar with other information search tools, such as catalogues, bibliographic databases, and discovery tools (Callicott & Vaughn, 2005; Gardner & Eng, 2005; Giustini, 2005; Levine-Clark & Kraus, 2007; Meier & Conkling, 2008; Bramer et al, 2013; Gehanno, Rollin & Darmoni, 2013; Stirbu et al, 2015; Breeding, 2015).

Beyond the analysis of Google Scholar as a search tool, its continued growth and the provision of citation counts that are not biased against the original source, language and format of the citing document, have led to a growing interest from the bibliometric and webometric community in studying this product as a tool for the evaluation of research activity (Torres-Salinas, Ruiz-Pérez & Delgado López-Cózar, 2009; Aguillo, 2011). This research has focused primarily on assessing the quality of the bibliographic and bibliometric data provided (Jacsó 2005; 2006; Bar-Ilan, 2010; Jacsó 2008a; 2008b; 2012) and its correlation with the indicators developed by Web of Science and Scopus (Bakkalbasi et al, 2006; Bar-Ilan, 2007; Bar-Ilan, Levene & Lin, 2007; Cabezas-Clavijo & Delgado López-Cózar, 2013; Harzing & Alakangas, 2016). Studies generally found significant correlations between the various data sources; however, not all studies found equally high correlations. Moreover, the fact that Google Scholar covers a far broader range of documents and that users cannot discriminate between the types of documents, compromised some of these comparative studies (Bornmann et al, 2009; Aguillo, 2012).

Other areas of research have focused on the usefulness of Google Scholar for obtaining unique citations (Yang & Meho, 2006; Meho & Yang 2007; Kousha & Thelwall, 2008; Bar-Ilan, 2010; Kousha, Thelwall & Rezaie, 2011; Harzing 2013; Orduña-Malea & Delgado López-Cózar, 2014) and, finally, on studying its coverage and its evolution over time (Aguillo, 2012; Khabsa & Giles, 2014; Harzing, 2014; Ortega, 2014; Winter, Zadpoor & Dodou, 2014; Orduna-Malea et al, 2015).

To date, empirical results have shown Google Scholar to be enormously useful when obtaining statistics on academic impact, especially for disciplines that use alternative channels of scholarly communication (in particular doctoral theses, books, book chapters, and conference proceedings), such as the Social Sciences, Humanities, and Engineering (Meho & Yang, 2007; Harzing & Van der Wal, 2008; Bar-Ilan, 2010; Kousha, Thelwall & Rezaie, 2011; Kousha & Thelwall, 2015, Martin-Martin et al, 2015). These are all disciplines in which the use of Google Scholar is deemed necessary to provide comprehensive information on academic impact. However, the literature also indicates that errors in the linking of citations and versions, and in the quality of the bibliographic data still persist, though to a lesser extent than in the early years (Winter, Zadpoor & Dodou, 2014; Orduna-Malea et al, 2015). This precludes Google Scholar's use as a standalone tool for scholarly assessment without prior filtering and processing of the data, an activity limited by the lack of options for the automated downloading of files.

Our review thus shows a significant accumulated knowledge base about Google Scholar as a search tool and a tool for evaluation of research activity, more specifically relating to its unique coverage when compared to other sources of publication and citation data. However, to the best of our knowledge there is no prior research on the capabilities of Google Scholar to identify highly-cited documents. In the context of Bibliometrics, highly-cited documents represent the most influential scientific works. Therefore, the identification of these documents allows detecting the most influential authors, topics, research methods, and sources of all times, and is thus a very important function of bibliometric research.

The identification of highly-cited documents in the Web of Science Core Collection (WoScc) or Scopus (the leading bibliographic databases for this purpose) is now quite a straightforward task. Both databases include, among the search sorting criteria (subject matter, date, author or journal), the number of times a paper is cited ("times cited" or "highest to lowest"). Therefore, it is simply a matter of selecting this option and the documents will be presented in descending order according to the number of citations received. Since there is no restriction in these databases on the number of documents that can be retrieved for a given query, identifying highly-cited papers is totally reliable. This enabled bibliometric studies of these documents to be conducted with ease. Conversely, the lack of a similar sorting feature in Google Scholar, together with the limitation of a maximum of 1000 results, i.e. document metadata, shown per query, raises the question of whether or not the identification of highly-cited papers is possible using this search engine.

Given the negative impact on the visibility of a document (and its authors) of not featuring in the top 1000 Google Scholar results for a specific query, search engine optimisation (SEO) is gaining popularity. This is an approach, already well-established in commercial environments (Ledford, 2009), whereby knowledge of the approximate sorting criteria of Google (a trade secret) has led experts to disentangle the key factors that influence the positioning of a website in the search results (Evans, 2007), one of which is the number of links that a website receives, a key indicator for webometrics (Orduña-Malea & Aguillo, 2014).

The application of these techniques to the academic environment (especially Google Scholar) has led to a new concept called Academic Search Engine Optimisation (ASEO). Beel, Gipp and Wilde (2010) define it as the creation, publication and modification of scholarly literature so as to facilitate crawling and indexing for the search engines, improving its subsequent position in the ranking. Although the number of citations seems to be one of the key indicators in this ranking process in Google Scholar (Beel & Gipp, 2009; 2010), other factors might positively or negatively influence the final rank that is achieved when a specific search is performed. We can distinguish between query factors (related to the nature of keywords), and document factors (language of articles, length of the articles, what words are used – or not used – in the title, abstract and keywords, or platforms in which documents are uploaded). In addition, the dynamism of the Web as well as the malfunction of some GS features (such as improperly linked versions) might affect the final rank as well.

These procedures (which may be artificially aimed at optimising the position of documents on the list of results to specific queries by authors) can be either a reflection of successful marketing and dissemination activities or, in contrast, the result of illicit activities designed to trick the search engines by manipulating certain data (Beel, Gipp & Wilde, 2010; Delgado López-Cózar, Robinson-García & Torres-Salinas, 2014). Existing ASEO procedures as well as the idiosyncratic way in which Google Scholar ranks results (which is kept under trade secret) mean that there is no *a priori* guarantee that users are able to retrieve all highly-cited documents. If this were the case, Google Scholar's subsequent usefulness as a tool for bibliometric evaluation would be severely limited.

Therefore, this paper has two main objectives:
- Verify whether it is possible to reliable identify the most highly-cited papers in Google Scholar, and indirectly,
- Empirically validate whether citations are the primary result-ordering criterion in Google Scholar for generic queries or whether other factors substantially influence the rank order.

## 2. Method

To accomplish the objectives formulated above, we propose first to construct generic queries (understood in this study to be a query that has not been filtered by author, journal or keywords) in order to minimize ASEO query factors, and then to calculate the correlation between the rank position achieved in the results and the number of citations the documents have received. A moderate-to-high correlation would indicate that citations are the key determinant for ordering results in Google Scholar and, as a result, we would be confident about the ability of Google Scholar to identify highly-cited documents.

We defined a generic query through conducting a null query (search box is left blank), filtering only by publication year using Google Scholar's advanced search function. In this way, we avoided the sampling bias caused by the keywords of a specific query and by other academic search engine optimisation issues. In order to work with a sufficiently large data sample, a longitudinal analysis was carried out by performing 64 generic null queries from 1950 to 2013 (one query per year). Whereas 2013 was the last complete available year when our data collection was carried out, 1950 was selected because this particular year reflected an increase in coverage in comparison to the preceding years (Orduna-Malea et al, 2015). After this, all the returned documents (a maximum of 1000 per query) were listed, obtaining a final set of 64,000 results.

This process was carried out twice (28 May and 2 June 2014). The first time, it was performed from a computer connected to a WoScc subscription via IP range to obtain WoScc data embedded in Google Scholar (http://wokinfo.com/googlescholar); the second time from a computer with a normal Internet connection. This functioned as a reliability check, as it allowed us to confirm that the two datasets contained the same records. After this reliability check, the html source code for each of the search engine results pages for each query was parsed and downloaded, and all bibliographic information for each result (taken only from the primary version of the document) was extracted (supplementary material available at https://dx.doi.org/10.6084/m9.figshare.1224314.v1). The available details for each bibliographic field are represented in Figure 1.

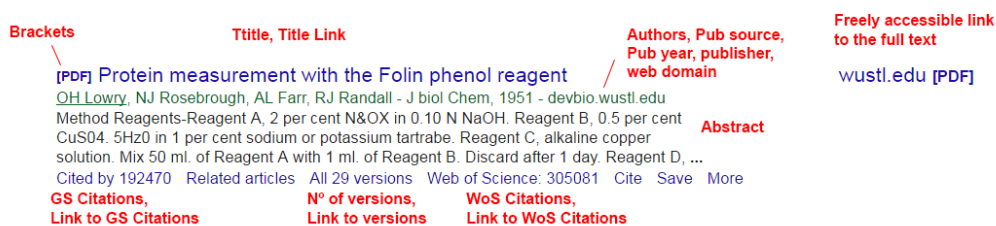

**Fig.1.** Google Scholar's bibliographic fields in the search engine results page.

Among the different information elements gathered, the following were processed in order to meet the objectives of this study (the remaining elements are unlikely to influence the ranking):

- **Rank**: position that each document occupies in the Google Scholar search engine results
- **GS Citations**: number of citations the document has received according to Google Scholar in the time the query was performed.
- **Number of versions**: number of versions GS has found of the documents.
- **Publication date**: year when the document was published.

Since Google Scholar doesn't provide information about the language of the documents, it was manually checked by observing WoScc data (when possible) as well as the language in which the title and abstract of the document were written.

All these data were then exported to a spreadsheet in order to be statistically analysed. Since citation data follow a skewed distribution, a Spearman correlation was calculated in order to find a relationship between citations and rank position.

## 3. Results

The overall correlation between the number of citations received by the 64,000 documents and the position they occupied on the results page of Google Scholar at the time of the query is $r= -0.67$ ($\alpha < 0.05$). Figure 2 shows the value of this correlation for each of the 64 years analysed. The aim of this year-by-year correlation analysis was both to investigate its possible evolution in time and to ascertain its value for the 1000 maximum results given for each query.

The average annual value of the correlation coefficient is very high (negative values for the correlation are due to position 1 being better than position 1000) and stable ($\bar{x}$= -0.895; $\sigma$= 0.025). In recent years (except 2013) the correlations are slightly lower than the average value; for example, the lowest values found correspond to 2006 and 2007. Even so, these correlation coefficients are still very high.

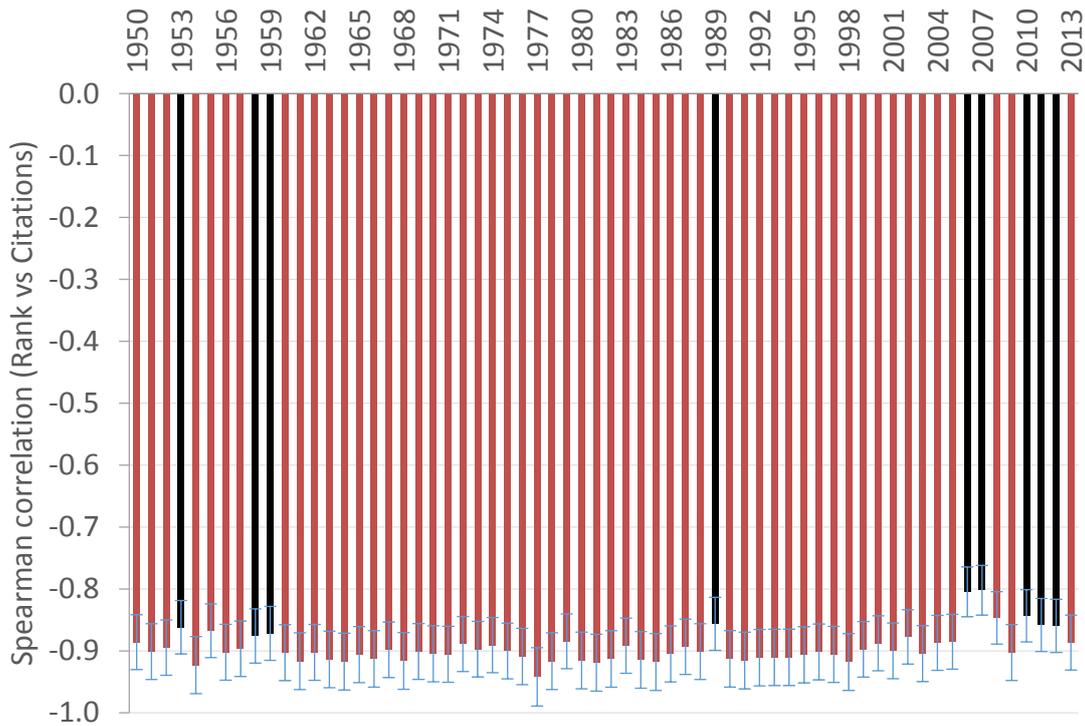

**Fig.2.** Spearman correlation between the number of citations received by documents in Google Scholar and the rank position they occupy in the search engine results page.

The fact that we obtained higher correlations in the annual samples than in the overall data indicates that there is a small deviation in the relationship between citations and a higher position in the list of results, which accumulates over 64 annual queries. In order to verify whether or not this deviation is concentrated in a specific area of the search engine results list, we proceeded to plot the dispersion between the position rank and citations received rank (Fig. 3), marking the observation points according to the language of the document (English in green; Not English in red).

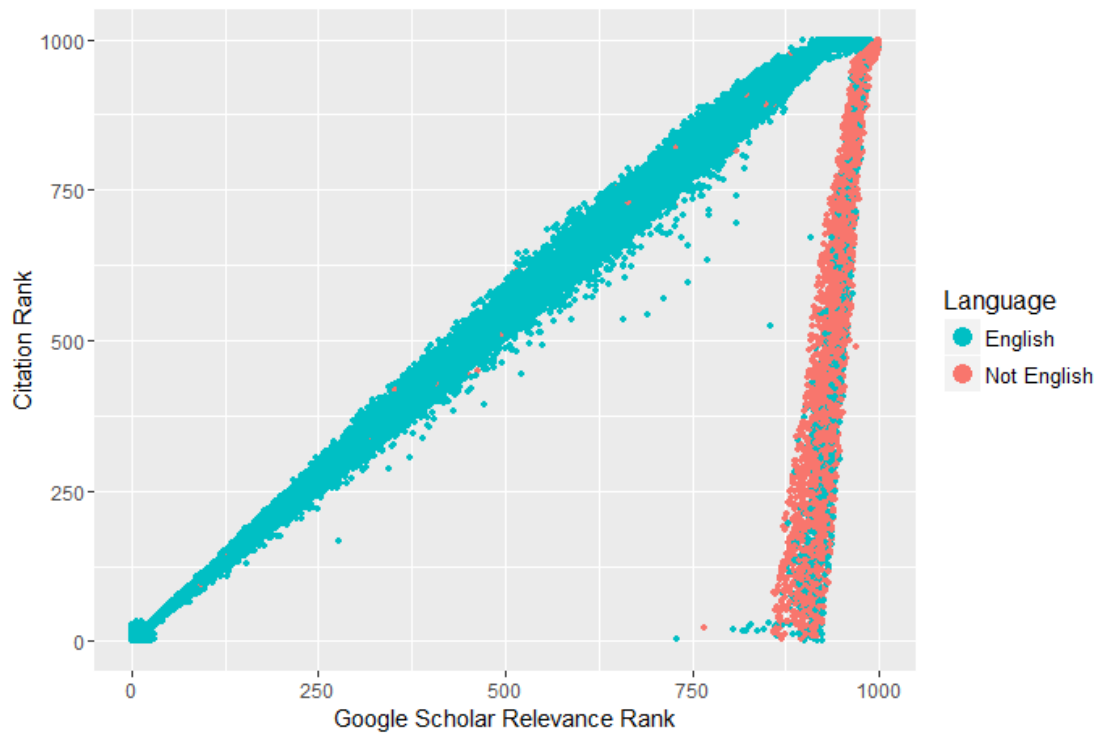

**Fig. 3.** Relationship between the number of citations of documents in Google Scholar and the rank position they occupy in the search engine results page.

In Figure 3, the results located in the first 900 positions of each search are displayed in green, while the results in the last 100 positions are shown in red. In this way we can see clearly how, until approximately the 900th position, the Google Scholar sorting criteria are based largely on the number of citations received by each result. However, after approximately the 900th position, the data show erratic results in terms of the correlation between citations and position.

The correlation for the results placed amongst the top 900 positions is r= 0.97 ($\alpha$ < 0.01). However, the correlation obtained for results in the last 100 positions is only r= 0.61 ($\alpha$ < 0.01). In this case we calculated the Pearson correlation, as discrete ranking positions were being compared for both variables. Although the sample size is different in the calculation of these correlations (900 versus 100), the data indicate the existence of unexpected results for the last 100 positions of the Google Scholar results page, i.e. some highly-cited documents are found in very low positions.

The positions occupied by the documents that received the highest number of citations in each year partly corroborate this irregular behaviour in Google Scholar. Figure 4 shows how in only 11 of the 64 years analysed (17.2%), the most-cited document that year is ranked first in the results, while in 32.8% of the years, this document is among the top three. However, sometimes the most highly cited document occupies a very low position. The most extreme case was detected in 1978, where the document with the largest number of citations appeared in the 917th position.

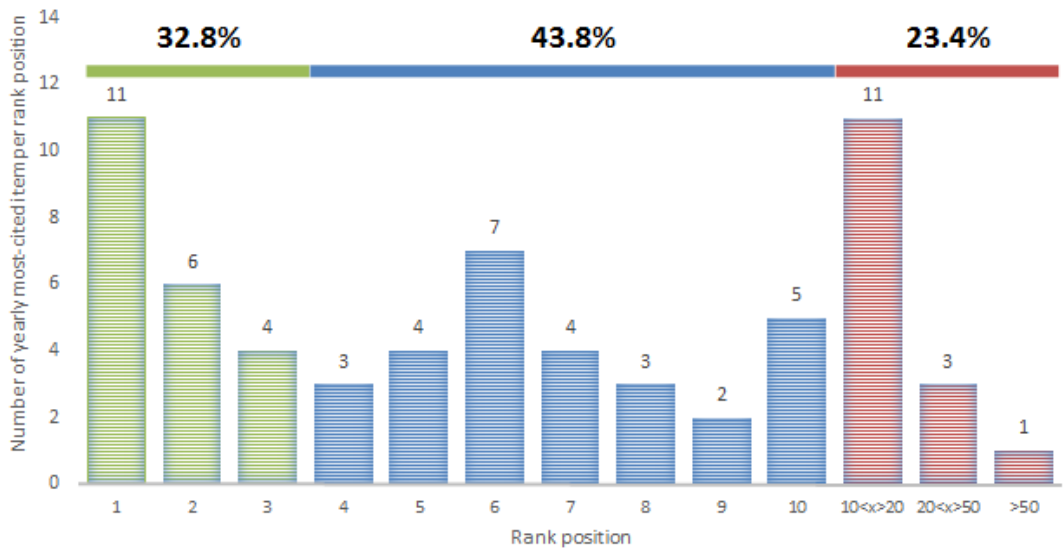

**Fig. 4.** Frequency of position occupied by yearly most-cited document in the search engine results pages.

At the other end of the scale, the document located in the last position (1000) is the document with the fewest citations in 60.9% of the years. In the years that this is not the case, the difference between the least-cited document and the lowest-ranked document is never greater than 10 citations; therefore, the number of citations received by the document located at position 1000 is a good reflection of the citation threshold level. Figure 5 shows precisely this threshold value per year: around 50 citations during the early years of analysis (1950-1960), subsequently climbing to over 200 citations between 1985 and 2000, and from then on falling to around 50 citations again in 2010. The rise of this threshold value (especially the last years of 20$^{th}$ century) is attributed to the increasing scientific output worldwide. However, the decline from 2000 onwards is unexpected. Though this will need to be empirically tested, we attribute the decrease of this threshold value on the increase of citations to old documents, a phenomenon in which the Google Scholar's rank is precisely contributing (Martin-Martin et al, 2016). In 2013, an atypical value (133 citations) was obtained. The likely reason for this issue will be discussed in the next section.

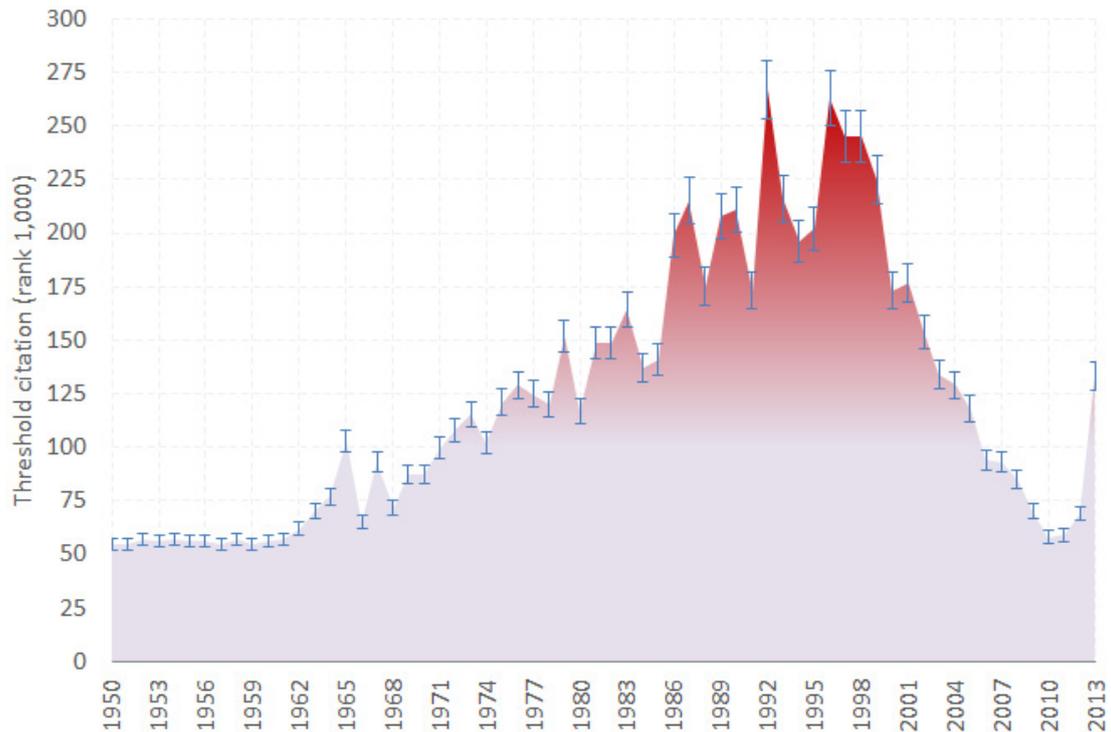

**Fig. 5.** Number of citations received per year by the document ranked 1000.

## 4. Discussion

Generic queries minimise the effect of those academic search engine optimisation practices that are influenced by query terms. Unfortunately, the relationship between the rank position of the results and the citations received by them may be influenced or determined by other external variables, such as the dynamism of the Web, the malfunction of some Google Scholar features, and ASEO practices determined from specific document characteristics. These are three aspects that all require detailed discussion.

### 4.1. Dynamic nature of the academic search engine

The way search engines (not only academic search engines, such as Google Scholar, but also general search engines, such as Google or Bing) function can cause two identical queries, made on different computers in different geographical locations, or simply repeated after a short period of time, to generate slightly different results (Wilkinson & Thelwall, 2013). This, in turn, can cause some documents to appear or disappear, or to move to another position within the search results page. Therefore, the results of a study like our current study should be considered from a general perspective, without entering too much into individual details.

Despite this, and in order to test the potential variability of the search engine, we again compiled the sample of 64,000 documents using the same procedure described in the methodology four months later (4 October 2014), and compared both samples. Table 1 shows the number of documents from the first sample that are not retrieved in the second sample, ordered by position interval.

**Table 1.**
Number of missing documents between the two samples of 64,000 highly-cited documents (May and October, 2014).

| Rank interval | Missing documents | % Total (n= 64000) | % Partial (n= 9402) |
|---|---|---|---|
| 001 – 100 | 402 | 0.6 | 4.3 |
| 101 – 200 | 340 | 0.5 | 3.6 |
| 201 – 300 | 319 | 0.5 | 3.4 |
| 301 – 400 | 373 | 0.6 | 4.0 |
| 401 – 500 | 450 | 0.7 | 4.8 |
| 501 – 600 | 588 | 0.9 | 6.3 |
| 601 – 700 | 778 | 1.2 | 8.3 |
| 701 – 800 | 1176 | 1.8 | 12.5 |
| 801 – 900 | 1802 | 2.8 | 19.2 |
| 901 – 1000 | 3174 | 5.0 | 33.8 |
| **TOTAL** | **9402** | **14.7** | **100** |

It is clear from Table 1 that accuracy diminishes the lower the position of the documents in the ranking. 14.7% of the 64,000 documents retrieved in the second sample (9402) are not found in the first. However of these, 65.4% (6152) are concentrated in the last 300 positions. This might have been influenced by the fact that the documents in these lower positions obtained similar or even identical values for the number of citations received. Hence even a change of one or two citations in the four-month lapse could lead a document to be included or excluded from the top-1000 results. Therefore, when considering highly-cited documents that occupy lower ranked positions results do need to be taken with a grain of salt. The same conclusion can be drawn from the low correlation coefficients obtained in Figure 3 (citation rank vs position rank) precisely for documents located in these lower ranked positions.

**4.2. Google Scholar malfunction 1: Rank position and number of versions**

Versions are a feature patented by Google Scholar (Verstak & Acharya, 2013) that enables all copies of the same document that are available online to be identified, and subsequently aggregated into a single result (adding up all the citations that each version may have received). We argue that the number of versions of a document could affect the relationship between citations and the position of a document mainly in two ways: multi-version effect (related to the number of versions; though it is not considered a malfunction, it is included in this section for expository clarity) and incorrect functioning effect (malfunction related to the version aggregation process).

The multi-version effect concerns the possibility of documents with a greater number of versions to appear in a higher position. Intuitively, one would expect documents with more versions to be dealing with important topics or written by outstanding researchers. This may explain why these documents are widely disseminated in several platforms. Accordingly, these documents are to be found more easily. Both their expected quality and wide discoverability may generate the multi-version documents to have more citations and, consequently, to climb in the Google Scholar's rank.

To determine the influence of the number of different versions on positioning, we calculated the dispersion between the number of versions of a document and its position on the results page (Fig. 6).

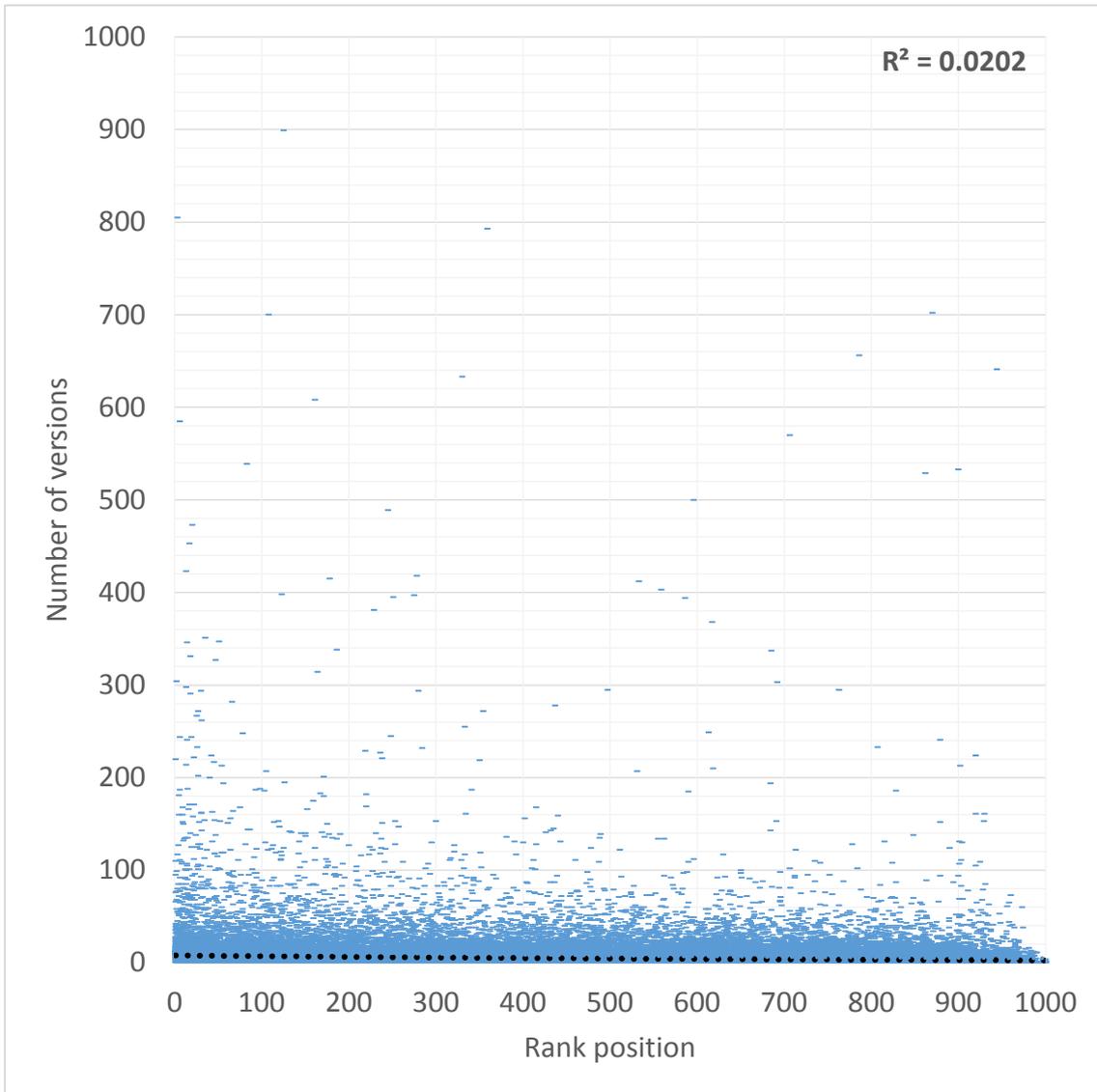

**Fig. 6.** Scatter plot of the number of versions and rank position values for the 64,000 documents in Google Scholar.

The correlation between the position of a document and the number of versions is low, but significant (r = -0.30; α < 0.01). The average correlation per year is slightly higher (r = -0.33; σ = 0.04). Figure 6 shows that, despite the wide dispersion of data, there is a slight concentration of documents with between 100 and 300 versions amongst the first 100 rank positions. In order to analyse this observation more precisely, Table 2 gives us the average number of versions of documents in a given year depending on their location in a range of positions. High average values (with equally high standard deviations) were identified in the documents in the first 100 result positions, although this behaviour does not follow any stable pattern, and there are some notable exceptions. Hence, there does seem to be a slight positive effect of the number of versions on the rank position of a document in the top 100 result positions.

**Table 2.**
Yearly average number of versions and standard deviation for documents according to rank position in Google Scholar (2014 – 2013).

| Rank position | Versions | | | | | | | | | | | | | | | | | | | |
|---|---|---|---|---|---|---|---|---|---|---|---|---|---|---|---|---|---|---|---|---|
| | 2004 | | 2005 | | 2006 | | 2007 | | 2008 | | 2009 | | 2010 | | 2011 | | 2012 | | 2013 | |
| | $\tilde{X}$ | σ | $\tilde{X}$ | σ | $\tilde{X}$ | σ | $\tilde{X}$ | σ | $\tilde{X}$ | σ | $\tilde{X}$ | σ | $\tilde{X}$ | σ | $\tilde{X}$ | σ | $\tilde{X}$ | σ | $\tilde{X}$ | σ |
| 1 - 100 | 23 | 35 | 19 | 20 | 25 | 30 | 23 | 28 | 22 | 50 | 18 | 25 | 23 | 63 | 16 | 20 | 16 | 25 | 12 | 11 |
| 101 - 200 | 22 | 25 | 23 | 26 | 21 | 20 | 26 | 34 | 19 | 37 | 17 | 21 | 13 | 13 | 18 | 70 | 12 | 34 | 15 | 44 |
| 201 - 300 | 16 | 19 | 28 | 52 | 24 | 32 | 28 | 59 | 19 | 24 | 16 | 20 | 20 | 57 | 12 | 12 | 11 | 24 | 8 | 12 |
| 301 - 400 | 18 | 21 | 22 | 27 | 17 | 16 | 15 | 11 | 21 | 78 | 15 | 19 | 17 | 26 | 12 | 13 | 10 | 27 | 8 | 9 |
| 401 - 500 | 16 | 19 | 16 | 14 | 18 | 18 | 18 | 16 | 17 | 19 | 17 | 32 | 17 | 20 | 11 | 9 | 9 | 17 | 7 | 7 |
| 501 - 600 | 13 | 11 | 16 | 14 | 14 | 15 | 16 | 13 | 17 | 16 | 13 | 11 | 16 | 40 | 10 | 12 | 8 | 12 | 9 | 23 |
| 601 - 700 | 15 | 16 | 18 | 14 | 17 | 15 | 15 | 11 | 15 | 18 | 16 | 21 | 16 | 16 | 10 | 12 | 9 | 19 | 6 | 6 |
| 701 - 800 | 14 | 11 | 14 | 11 | 16 | 17 | 15 | 12 | 13 | 9 | 16 | 17 | 12 | 10 | 9 | 11 | 8 | 9 | 12 | 65 |
| 801 - 900 | 12 | 10 | 15 | 15 | 11 | 17 | 17 | 54 | 10 | 9 | 11 | 12 | 11 | 9 | 11 | 25 | 20 | 87 | 8 | 19 |
| 901 -1000 | 10 | 64 | 5 | 15 | 3 | 5 | 3 | 6 | 4 | 13 | 7 | 23 | 7 | 22 | 5 | 9 | 4 | 5 | 6 | 13 |

In red: rank position for which the highest yearly average number of versions is obtained.

With regard to the incorrect functioning effect, we distinguish the following shortcomings:

a) Incorrect functioning leading to the omission of citations: the incorrect functioning of the version aggregation process could cause legitimate citations to a document to be omitted, causing it potentially to be excluded from the first 1000 results.
b) Incorrect functioning leading to the overestimation of citations: the incorrect functioning of the version aggregation process could cause citations to be wrongly attributed, thereby causing the document to be unjustly positioned amongst the top 1000 results.

However, given the low overall correlation detected between the number of versions and the position in the results, and without considering the exact position that each document should occupy if all existing versions were linked properly (which would require a systematic study focusing on this issue), we argue that version aggregation does not seem to affect greatly whether a document is included or excluded in the top 1000 results, i.e. the main objective of this study.

**4.3. Google Scholar malfunction 2: Rank position and publication date**

As was demonstrated in earlier studies, the results for Google Scholar's advanced option searches in a specific year (custom range) are not always entirely accurate (Orduna-Malea et al, 2015). This could mean that the actual year of publication of a document does not correspond with the year specified in the corresponding query. If this happens, a document may not appear among the results of a generic query (if it has no publication date) or it may appear among the results for another year (if it has an erroneous publication date). To determine the potential impact of this problem for the objectives of this study we performed two consistency tests (internal and external).

The internal consistency test verified whether the date of publication provided for each document corresponds to the date indicated in the advanced search for each of the 64 queries. The results of the test indicate that only in 2 documents (out of the 64,000 analysed) did the publication date not coincide with the date of the query. We can therefore conclude that the system works accurately at the technical level.

Another, quite different, issue is whether or not the date of publication provided by Google Scholar is correct. To this end, we conducted an external consistency test to cross-check the publication date of each document with the date provided by a controlled source independent of Google Scholar (in this case WoS). This is obviously assuming that WoS will provide correct data most of the time, though this is not always guaranteed, due to sporadic errors with online-first articles and other bibliographic data (Franceschini, Maisano & Mastrogiacomo, 2016).

We obtained a sample of the 64,000 documents (those that were linked to a WoS result), comprising 51% of the documents (32,680). The results of this process showed a match between the dates provided by both sources in 96.7% of the documents. Figure 7 displays the annual distribution of the documents in which there is no such match.

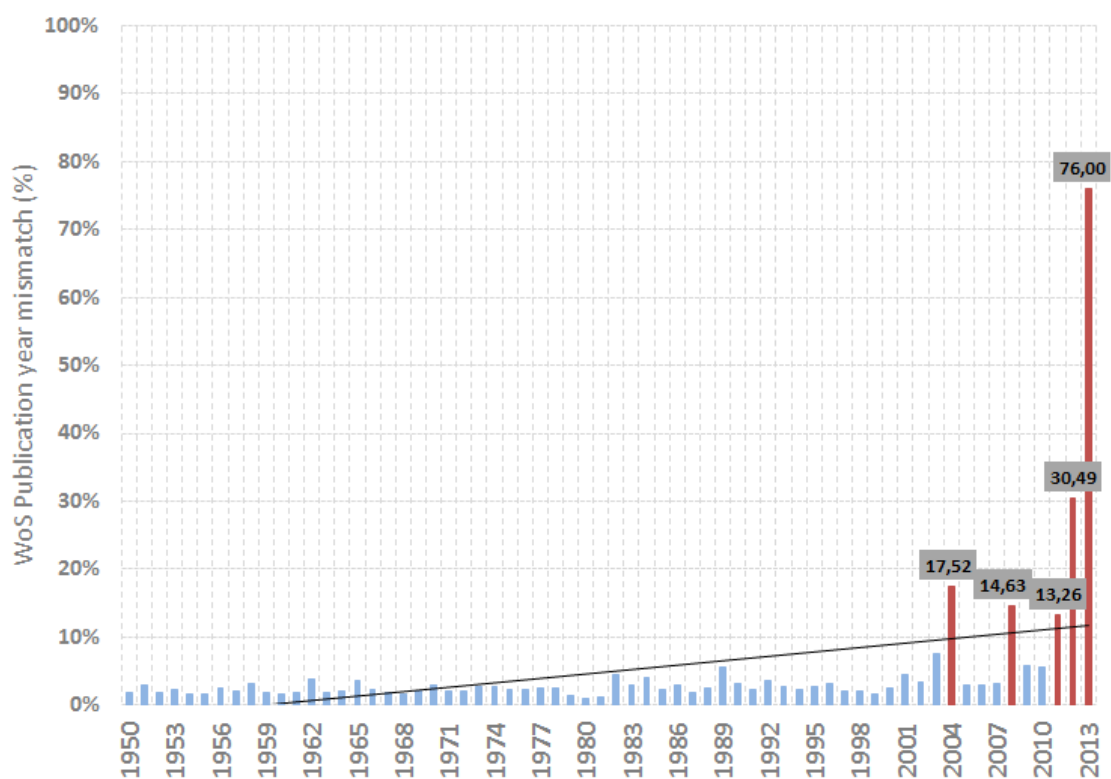

**Fig. 7.** Percentage of publication year mismatches between documents in Google Scholar and the Web of Science between 1950 and 2013.

As shown in Figure 7, there is a concentration of errors in recent years, especially over the last two years of the period analysed (2012 and 2013), in which the error rate shoots up (30.49% and 76% respectively). This could explain the atypical value previously shown in Figure 5 for 2013.

A detailed analysis of 2013 (Fig. 8) shows us how, out of the 19 errors detected this year, for 11 of them (58%) the error (difference between the year recorded by both sources) is more than 20 years, while only on two occasions is the error less than 2 years. These results therefore mean that this error cannot be attributed to the publication of preprints and/or periods during which the journal was under embargo. The Google Scholar practice of selecting the latest edition of a monograph as the main version seems to be the primary cause of these errors. Consequently, the small number of documents analysed for these two years (82 and 25 respectively) leads to a high proportion of mismatches. All different editions of a book (with their corresponding years of publication) are treated as versions by Google Scholar, after which Google Scholar selects as the primary version (the version used in this study) the document with the most recent publication date.

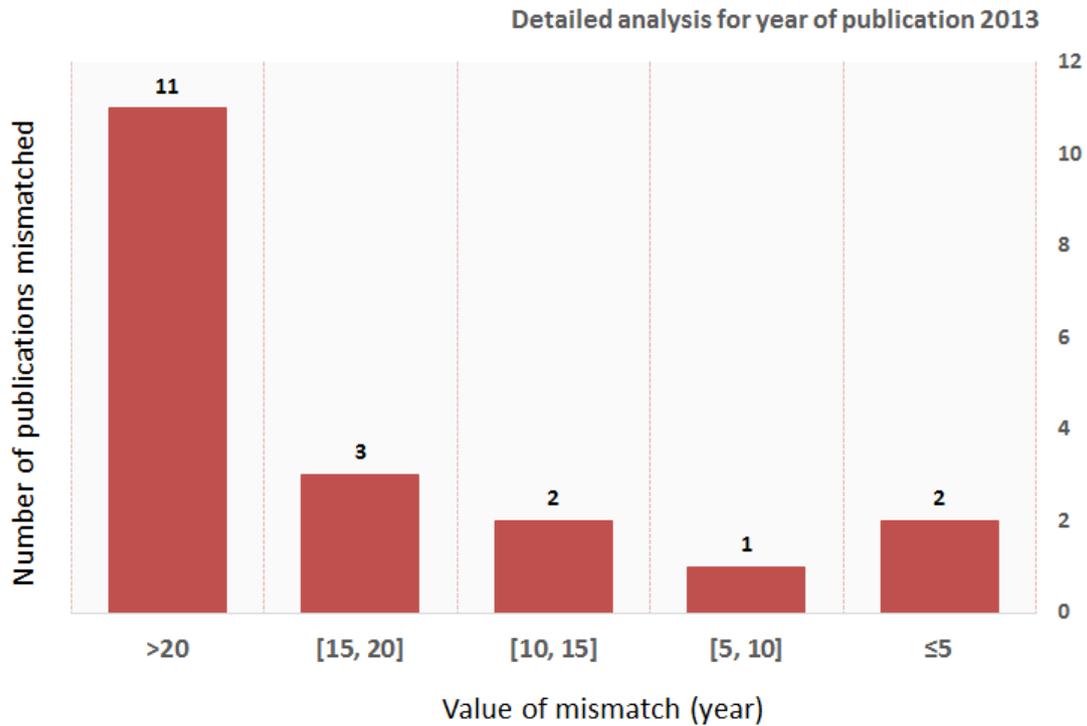

**Fig. 8.** Publication year mismatches between documents in Google Scholar and the Web of Science (2013).

However, the percentage error for the data sample as a whole is very small. If we add to this the fact that an error in the date can cause the document to appear in the wrong year, but not exclude it from the results of a generic search, we may safely say that the publication date does not significantly affect the ability of Google Scholar to identify highly-cited documents.

*ASEO document factor: Rank position and language of publication*

Finally, we looked at the possible influence of the language of publication on the rank position in the results. To study this effect in detail, we have analysed the percentage of documents published in English in the first and last 100 results of each year (Fig. 9).

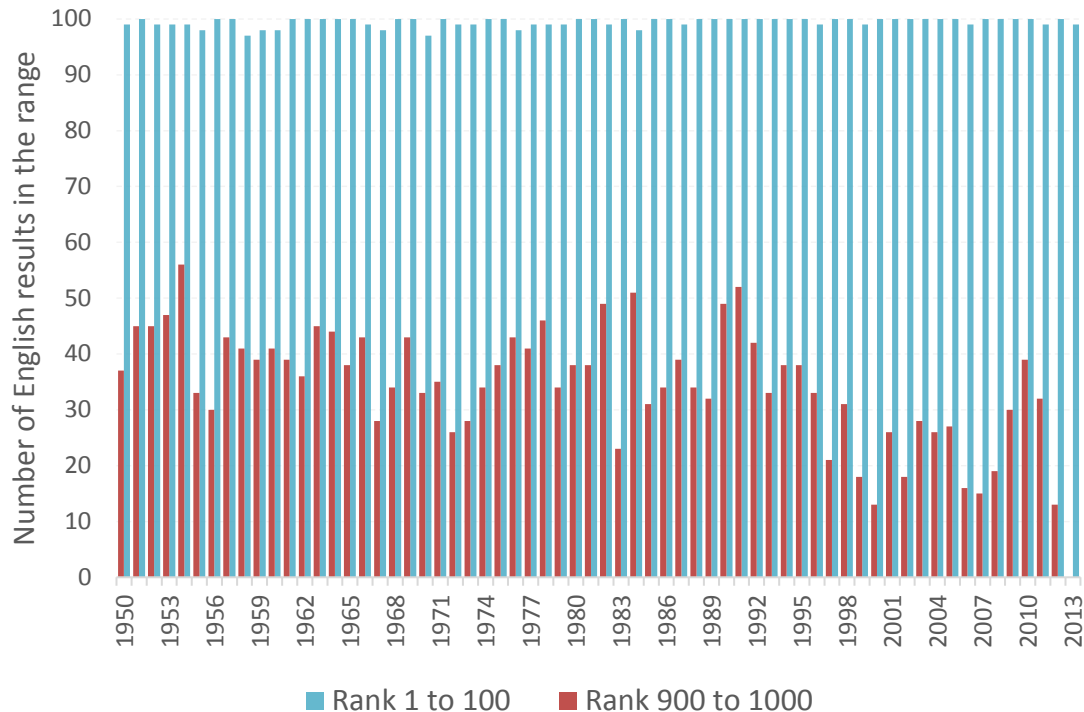

**Fig. 9.** Yearly percentage of English documents among the first and last 100 ranked documents.

The annual average number of documents in English for results within the first 100 positions is 99.5. Therefore, the presence of documents in other languages within this range is abnormal. When analysing this same percentage for the documents in the last 100 positions, the results change significantly. The annual average drops to 34.2%.

The high presence of documents in languages other than English (often with very high levels of citations) in the last 100 positions could help explain the low correlations identified in this range between the ranking positions and citations received (Fig. 3). This may have been due our choice of interface language (English was selected during the study). It should be noticed that users cannot select the actual language of documents in the Google Scholar's advanced search features but instead select the language of the website. The latter is primarily identified by detecting the geographic domain in which the document is available online (for example, .nl, .es or .jp) and does not guarantee the website is actually in that language. This may explain the fact that some documents written in English but with their primary version hosted in non-Anglophone countries' web domains do appear in lower positions in spite of receiving a large number of citations.

Therefore, if the queries had been conducted by restricting the geographic web domains to Anglophone countries, it is likely that the correlation coefficients would have been significantly higher, especially in the last quartile of the ranking results. However, this obviously would result in a biased interpretation of which publications are most highly cited, limiting the results largely to English-language publications. The effect of the choice of interface language on the results has already been partly studied in the past (Lewandowski, 2008). However, this effect should be tested empirically, and methodically, in the future to assess the impact of the interface language with greater accuracy.

## 5. Conclusions

A significant and high correlation between the number of citations and the ranking of the documents retrieved by Google Scholar was obtained for a generic query filtered only by year. The fact that we minimised the effects of academic search engine optimisation, together with the size of the sample analysed (64,000 documents), leads us to conclude that the number of citations is a key factor in the ranking of the results and, therefore, that Google Scholar is able to identify highly-cited papers effectively. Given the unique coverage of Google Scholar (no restrictions on document type and source), this makes it an invaluable tool for bibliometric analysis.

However, the correlation that was obtained, though high, was not excellent because of external factors (especially the language of publication and the geographic web domain where the primary version was hosted) that mainly affected the results at the bottom of the list (approximately the last 100). Restricting the language of the results to match the interface language may help to improve accuracy in the search for highly-cited papers, although this obliges us to perform as many queries as the languages we wish to analyse. Unfortunately, users can only restrict the language of the website, and this procedure is far from optimal as it mainly relies on geographic web domains.

Other factors, such as the date of publication (when erroneous) or the number of versions (multi-version effect and incorrect functioning of version aggregation effect) only have an incidental impact, and do not compromise the proven ability of Google Scholar to search for highly-cited documents.

Therefore, we conclude that Google Scholar can be used to reliably identify the most highly-cited academic documents. Given its wide and varied coverage, Google Scholar has become a useful complementary tool for Bibliometrics research concerned with the identification of the most influential scientific works.

## Acknowledgements

Alberto Martín-Martín enjoys a four-year doctoral fellowship (FPU2013/05863) granted by the Ministerio de Educación, Cultura, y Deportes (Spain). Enrique Orduna-Malea holds a postdoctoral fellowship (PAID-10-14), from the Polytechnic University of Valencia (Spain).